\documentclass[draft]{agujournal2018}

\usepackage{apacite}
\usepackage{url} 
\usepackage{comment}
\draftfalse

\journalname{JGR-Space Physics}

\begin{document}

%
%


\title{Spatially Resolved Observations of Meteor Radio Afterglows with the OVRO-LWA}

%
%




\authors{S. S. Varghese\affil{1}, J. Dowell\affil{2}, K. S. Obenberger\affil{3}, G. B. Taylor\affil{2}, M. Anderson\affil{4},   and G. Hallinan\affil{5}}


\affiliation{1}{SETI Institute, Mountain View, CA, USA}
\affiliation{2}{University of New Mexico, Albuquerque, NM, USA}
\affiliation{3}{Air Force Research Laboratory, Kirtland AFB, Albuquerque, NM, USA}
\affiliation{4}{Jet Propulsion Laboratory, Pasadena, CA, USA}
\affiliation{5}{California Institute of Technology, Pasadena, CA, USA}





\correspondingauthor{Savin Shynu Varghese}{svarghese@seti.org, savinshynu@gmail.com}




\begin{keypoints}

\item An all-sky imaging transient search with the OVRO-LWA detected 5 meteor radio afterglows (MRAs).

\item The radio emission was resolved through imaging and their spectral index maps were created.

\item The unpolarized nature of the resolved emission favors resonant transition radiation as the emission mechanism of MRAs.


\end{keypoints}

%
%

\justify
\begin{abstract}
We conducted an all-sky imaging transient search with the Owens Valley Radio Observatory Long Wavelength Array (OVRO-LWA) data collected during the Perseid meteor shower in 2018. The data collection during the meteor shower was motivated to conduct a search for intrinsic radio emission from meteors below 60 MHz known as the meteor radio afterglows (MRAs). The data collected were calibrated and imaged using the core array to obtain lower angular resolution images of the sky. These images were input to a pre-existing LWA transient search pipeline to search for MRAs as well as cosmic radio transients. This search detected 5 MRAs and did not find any cosmic transients. We further conducted peeling of bright sources, near-field correction, visibility differencing and higher angular resolution imaging using the full array for these 5 MRAs. These higher angular resolution images were used to study their plasma emission  structures and monitor their evolution as a function of frequency and time. With higher angular resolution imaging, we resolved the radio emission size scales to less than 1 km physical size at 100 km heights. The spectral index mapping of one of the long duration event showed signs of diffusion of plasma within the meteor trails. The unpolarized emission from the resolved radio components suggest resonant transition radiation as the possible radiation mechanism of MRAs.

\end{abstract}

\section{Introduction} \label{intro}
Transient searches below 100 MHz using the all-sky imaging capabilities of the Long Wavelength Array (LWA) stations in New Mexico (LWA1; \citet{tay12} and LWA-SV; \citet{cra17})  resulted in the first detection of intrinsic radio emission from meteors known as meteor radio afterglows (MRAs; \citet{ob14b}). Several observational campaigns were conducted to further understand the properties and origin of the radio emission \citep{ob15a,ob15b,ob16a, ob16b,var19b,var21}.  Currently, we know that the emission is non-thermal, unpolarized, broadband between at least 20-60 MHz and observed to have an occurrence cutoff below 90 km altitude \citep{ob14b,ob15a,ob15b,ob16b,var21}.  The study conducted by \citet{ob20} demonstrated that the MRA emission is spatially and temporally associated with the persistent trains from bright meteors. Furthermore, \citet{var21} utilized the broadband imager at LWA-SV to understand the spectral distribution of MRAs. In this study, the spectral indices of 86 MRAs peaked at -1.73.
Even though the spectral parameters were not correlated with the physical properties of MRAs, the duration of MRAs were found to correlate with the local time of occurrence, luminosity of MRAs, incidence angle and kinetic energy of the parent meteoroid.

The radio and optical observations to date have been able to improve our understanding of the MRA emission. However, an estimate of the emission size scales and an insight into the radiation mechanism is yet to be fully understood. The two existing hypotheses for the radiation mechanism of MRAs are the electromagnetic conversion of the electrostatic plasma oscillations within the plasma trail (Langmuir waves) and the resonant transition radiation (RTR; \citet{plat02}), which has been discussed in \citet{ob20} and \citet{var21} in regards to MRAs.

RTR is a process similar to Cherenkov radiation, where hot electrons moving through a turbulent plasma can radiate due to the changing refractive index with position. \citet{ob20} showed that the hot electrons for such a process could be produced by oxidation of negatively charged metallic species. RTR is the better match for the observations  because it predicts a broad spectrum of emission, no polarization and a relatively isotropic radiation pattern. Langmuir wave conversion on the other hand should be a coherent process with intrinsic polarization. The radiation pattern would depend on the size scale of the coherent region. For RTR,  we would expect the MRA luminosity and spectra to be related to both the level of turbulence and rate of oxidation. The oxidation rate is likely also related to the level of turbulence, where more turbulence would be associated with more mixing of meteoritic metals and atmospheric oxygen.

\citet{var19b} calculated the luminosities of 32 co-observed MRAs using the LWA1 and LWA-SV stations and showed that the MRAs radiate isotropically.  This suggested two possibilities for the size scales of MRA emission. First is an incoherent emission from large scale regions in the plasma trail which would be non-directional and unpolarized. Second is an incoherent addition of different coherently emitting regions within the meteor plasma. In that case, each individual coherent regions could be polarized and directional. However, they could appear as unpolarized and isotropic when the emission is averaged over large scales. The first scenario is consistent with the RTR hypothesis and the second scenario is consistent with the Langmuir wave hypothesis.

The LWA1 and LWA-SV stations each have 256 dipole antennas arranged in an elliptical layout with a maximum baseline of 110 m in the N-S direction. This arrangement produces good ($u,v$) coverage (sampling for imaging) on short time scales. This makes a LWA station an ideal instrument for making  snapshot images  (5 seconds duration) of the sky with minimal side lobes from bright sources. 
One of the downsides of using  a single LWA station for imaging is the limited angular resolution preventing the precise localization and limiting our ability to resolve the spatial structure of radio sources. The limited angular resolution also increases the confusion noise in the image.

In radio interferometry,  the angular resolution of a telescope at a given wavelength is inversely proportional to the maximum separation between two antennas (maximum baseline). The longer the maximum baseline, the better the angular resolution of the telescope. The LWA1 and LWA-SV has maximum baseline lengths up to 110 m which limits the angular resolution to 4 degrees at 40 MHz. This angular distance translates to a $\sim$8 km physical lengths at 100 km heights and at a zenith angle of 30 degrees. Thus an 8 km long meteor plasma trail will appear as an unresolved point source within the LWA synthesized beam. In that case, we could either be observing radio emission from a 8 km scale plasma structure or an addition of several smaller resolved plasma regions within the LWA synthesized beam. Therefore, in order to better understand the size scales of the MRA emission regions, higher angular resolution observations are required. Under the RTR hypothesis, increased resolution should give an enhanced picture of the turbulent structure along the trail.

The stage II of the Owens Valley Radio Observatory LWA (OVRO-LWA) telescope located in California has a maximum baseline length of 1.5 km.  This provides  $\approx$18 arcminute angular resolution at 40 MHz.  The differences in the data capture and processing pipelines limits OVRO-LWA from continuous all-sky monitoring for transient sources. Instead, data collected in observational campaigns over a few days are used for transient searches.

\citet{mar19} conducted the first OVRO-LWA transient survey utilizing 31 hours of all-sky images. The survey did not detect any cosmic or atmospheric transients on time scales of 13 s, 39 s, 2 minutes and 6 minutes. The cosmic transient detection by \citet{var19a} utilized 10,240 hours of all-sky images from the two LWA stations (LWA1 and LWA-SV). \citet{ob14b} reported an average detection of 60 MRAs per year where most of the events peak during the time of meteor showers. The 31 hour transient search time scale with OVRO-LWA  is much shorter compared to the New Mexico LWA transient search timescales for the detection of cosmic transients and MRAs. Also, the 31 hour survey conducted between 17 - 18,  February 2017 did not coincide with any of the known meteor showers \citep{mar19}. The shorter duration of the all-sky observations and date of observations not overlapping with the known meteor shower could have been the reasons OVRO-LWA did not detect any MRAs which are frequently observed by the LWA stations in New Mexico. Also, an aggressive radio frequency interference (RFI) flagging in these earlier OVRO-LWA searches might have masked times when meteor scatter appeared, even though MRA emission could have been present. Similarly, the transient search with OVRO-LWA conducted by \citet{hua22} utilized 137 hours of observation between 21 - 26, March, 2018. This search focused on the detection of the LOFAR ILT J225347+862146 transient \citep{ste15} and the observational dates did not overlap with any of the known meteor showers in the Northern hemisphere.

Recent work conducted by \citet{tam21} demonstrated the first detection of MRAs outside of New Mexico with a radio telescope other than the LWA. This study utilized simultaneous radio and optical observations with the LOFAR/AARTFAAC12 radio telescope \citep{van13, pr16} and the CAMS BeNeLux low-light video network \citep{jen11}. The study detected broadband radio emission between 30--60 MHz associated with many persistent trains. However, the study integrated across the entire band, prevented discrimination between the forward scatter from RFI and actual broadband MRAs. Even though the radio observations were conducted using LOFAR/ARRTFAAC12 with maximum baseline of 1.2 km, the study did not report size scales of the emission regions. At the same time, this work with OVRO-LWA utilizes baseline lengths up to 1.5 km, can provide better spatial resolution and we intend to study the emission size scale of MRAs.

In this work, we  have utilized 4 days of data collected from OVRO-LWA during the Perseids meteor shower in 2018 and report the detection of 5 MRAs at higher angular resolution. The higher angular resolution images are corrected for near-fields effects and their spectral index maps are created to understand the evolution of meteor trails.

\section{Observations} \label{obs}
The OVRO-LWA is a low frequency radio telescope operating between 27-84 MHz located at the Owens Valley, California \citep{mich18, mar19}. The array utilizes similar LWA antennas used at the LWA stations in New Mexico \cite{tay12, cra17}. The development of OVRO-LWA took place in three distinct stages. The stage II of the array has 251 dipole antennas within the 200 m core and 32 antennas spread across the NW direction of the core with baseline lengths up to 1.5 km. For this work, data from the stage II of the array between 30--50 MHz were collected in 4 subbands: 30.0--32.6 MHz, 35.2--37.8 MHz,  40.4--43.0 MHz and 45.6--48.2 MHz. Each subband has a bandwidth of 2.6 MHz providing a total 10.4 MHz of bandwidth across the frequency range of 30-50 MHz. The data were collected during the peak of Perseid meteor shower on 10, 11, 15  and 16 Aug, 2018.  Due to the technical difficulties, the telescope could not observe from Aug 11, UTC 19:00 to Aug 15, UTC 21:00,  resulting in a total of 69 hours of observation across 4 days. The voltages collected from the dipoles are sent to the Large-Aperture Experiment to Detect the Dark Age (LEDA) correlator \citep{koc15}, where the signals from each pair of antennas are cross correlated and integrated down to 13 s data sets. This correlated data product is generally known as complex visibilities in radio astronomy. The raw visibility data sets for each 13 s are later converted to a Common Astronomy Software Applications (CASA; \citet{mc07}) measurement set (MS) format for calibration and imaging purposes.
The MS format contains all the visibility data and metadata of the observations. The observation campaign collected $\sim$ 8.6 terabytes of data in the MS formats which were copied over to the LWA data archive for further analysis.

\section{Analysis}\label{ana}
\subsection{Calibration} \label{cal}
The calibration and imaging of the measurement sets were conducted with CASA \citep{mc07} and mostly followed the procedures outlined in \citet{mar19}. In the initial step of pre-processing, \texttt{flagdata} task in CASA was used to flag 20\% of the antennas that showed with signal loss or distorted bandpasses and 1\% of baselines to reduce the effects of cross-coupling from adjacent signal paths. This step also removed bad 
channels affected by RFI.  Next, a bandpass calibration of the data was performed using a simplified two point source model. This model consists of  two bright sources, the radio galaxy Cygnus A (Cyg A) and the supernova remnant Cassiopeia A (Cas A) with their flux densities and spectral indices adopted from \citet{bar77}.  The CASA \texttt{bandpass} task was used to derive antenna gains per channel using the two point source models. In order to to reduce the effects of diffuse Galactic emission,  solutions were derived using only baselines longer than 15 $\lambda$.  The bandpass calibration for a day is derived using a single measurement set at a time when Cyg A transits through the highest elevation in the sky. The derived calibration solutions are generally stable for 24 hours and are used to calibrate rest of the datasets from the same day. For more details on calibration, see, \citet{mar19}.

\subsection{Lower Angular Resolution Imaging} \label{lowim}
In order to detect MRAs, we initially used the core array for imaging to achieve an angular resolution similar to the LWA stations in New Mexico. This method is best for finding  sources in the near-field such as MRAs. The transition from near-field to far-field happens when,
\begin{equation}
    d > \frac{2D^{2}}{\lambda},    
\end{equation}
where $d$ is the distance to the source, $D$ is the diameter of the antenna/array and $\lambda$ is the wavelength.
 Most radio telescopes focus to infinity and the correlator assume that the incoming waves are plane waves.  For an array with maximum baseline of 200 m, this transition happens at 8 km when observed at 30 MHz. However for an array with baseline length of 1.5 km, the transition occurs at 450 km while observing at 30 MHz. Therefore, the plane wave approximation will not hold for MRAs occurring at 100 km heights when observed with the full array. For the detection of MRAs, the near-field complications can be avoided by using the core array with a maximum baseline of 200 m. This provides a maximum resolution of $\sim$ 2 degrees at 40 MHz. These lower angular resolution images can be searched for MRAs. Once we find an MRA, then the higher angular resolution imaging correcting for the near-field effects can be carried out to study the small scale emission regions associated with the source.  As the angular resolution decreases, the confusion noise in the image increases, which decreases the separability of sources within a synthesized beam. At the same time, the mutual coupling between nearby antennas changes the primary beam pattern of each antenna element and introduces a strong sidelobe pattern associated with bright sources. In order to produce good images, we need to peel out the bright sources and their associated sidelobe patterns. At LWA frequencies with the 200 m core array, the confusion noise is greater than the noise pattern introduced by mutual coupling, and peeling of sources are not required for transient detection. More details on peeling of bright sources are discussed in Section \ref{sec:peel}.
 
 Due to computational limitations, we only imaged the first subband (30.0 - 32.6 MHz) of the data for detection. The imaging was conducted in CASA using the task \texttt{clean} in the channelized imaging mode, with an image size of 256$\times$256 pixels, a cell size of 30 arcminute, ($u,v$) range restricted to less than $30\lambda$ to use the core array, a natural weighting scheme for higher surface brightness and without any deconvolution iterations. Once, we find an interesting broadband source in one subband, then the rest of the subbands were imaged to obtain a higher SNR image. The correlator produced roughly 6646 measurement sets for a day and imaging them sequentially required several weeks. In order to carry out faster imaging, we utilized the High Performance Computing (HPC) facility at the Center for Advanced Research Computing (CARC), University of New Mexico. The HPC tool, GNU Parallel software \citep{tan11} along with CASA was used to parallellize the calibration and imaging of the measurement sets. After calibration and imaging, the images produced in an hour were converted to a single Hierarchical Data Format (HDF5; \citet{hdf5}) file containing headers and data. The HDF5 files produced in a day are later sent to the transient search pipeline for detection of MRAs.

\subsection{Transient Search Pipeline} \label{pipe}
The all-sky images produced in a day are input to the transient search pipeline for finding transient sources. The transient search is conducted on the timescales of 13, 26 and 65 s. Here, we have utilized the transient search and candidate filtering pipeline described in \citet{var21} for the LWA-SV station. Briefly, the pipeline uses an image subtraction algorithm in which an average of previous images within the last 26-65 seconds (depending on search timescale) is subtracted from a running image and pixel regions greater than 6 sigma flux detection are marked as transients candidates. Bright steady sources were masked to reduce scintillation of sources caused by the ionosphere. Similarly, the sky below 25 degree elevation was masked to reduce false positives from RFI originating from sources near the horizon (airplanes, power lines, etc.).
The right ascension, declination and the time  of each event from the subtracted images were output as a transient list. These events are then sent to a Python script which can filter out low signal to noise ratio (SNR) events from their light curves. Once a source with good SNR is found, the spectral information of the source is used to rule out narrowband RFI sources. Also, slowly moving objects like airplanes are filtered out by the algorithm. Finally, a visual inspection of the filtered candidates is carried out to confirm the detection.

\subsection{Detection of MRAs}\label{mras}
\begin{figure}[h!]
	\centering
	\includegraphics[height = 9 cm, width = 12cm]{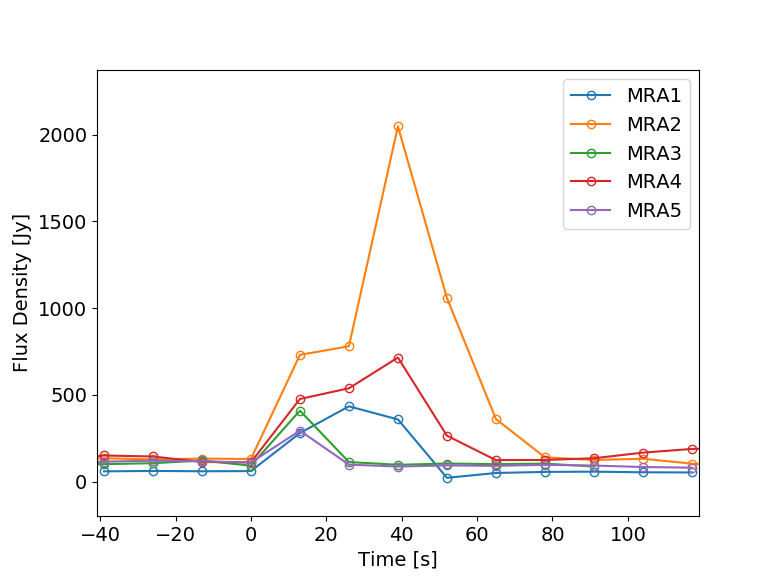}	
	\caption{Stoke I light curves of 5 MRAs. All the light curves are aligned with time = 0 for easier comparison.}
	\label{lc_multi}
\end{figure}

\begin{figure}[h!]
	\centering
	\includegraphics[height = 9 cm, width = 16cm]{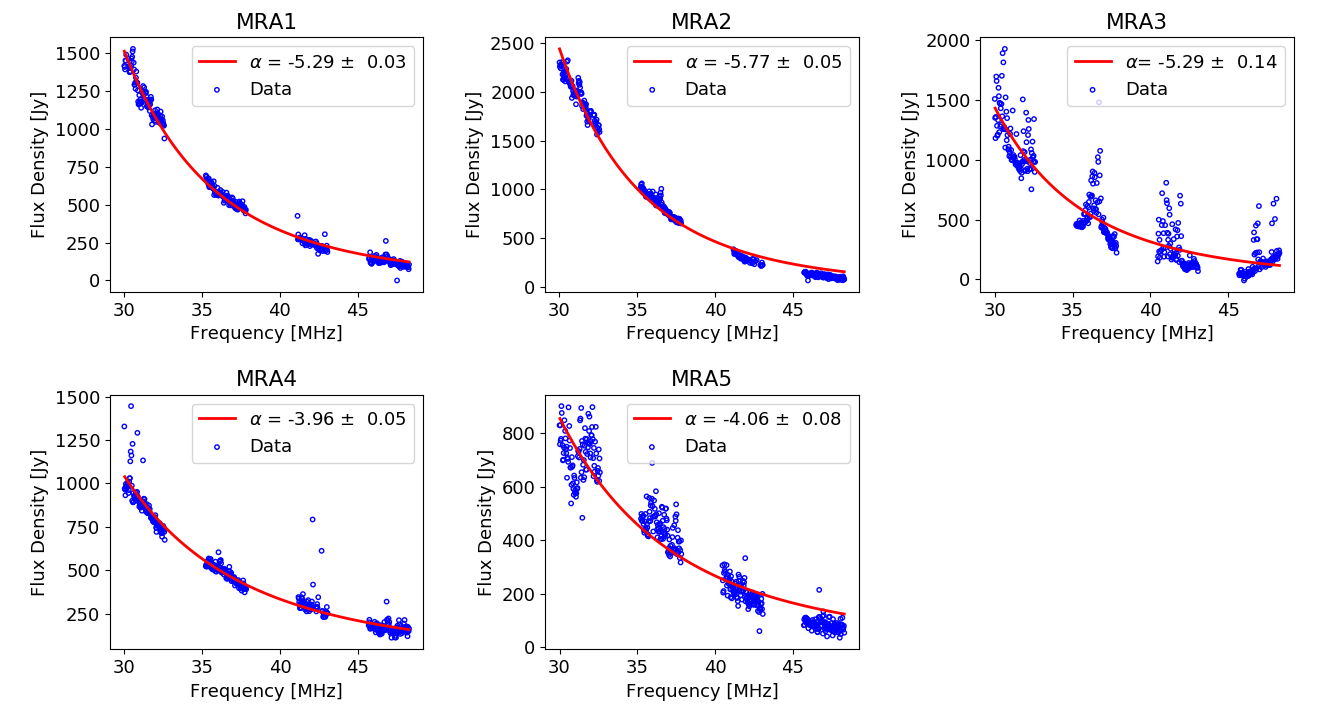}	
	\caption{Spectra of five MRAs fitted with the power law. The spectral index values from the fit are marked in each subplot.}  
	\label{spec_multi}
\end{figure}

The search resulted in the detection of 5 MRAs and their naming conventions, elevation angle, date and time of detection are listed in Table \ref{mra_table}. Figure \ref{lc_multi} shows the Stokes I light curves of all the detected MRAs. MRA3 and MRA5 are single integration (13 seconds) events whereas the rest of MRAs have longer durations. MRA2 is the longest event with 65 seconds duration. Figure \ref{spec_multi} shows the spectra of all MRAs at their peak flux density integration time fitted with a power law of the form, 
\begin{equation}
    S_{\nu} \propto \nu^{\alpha},
\end{equation}
where $S_{\nu}$ is the flux density, $\nu$ is the frequency and $\alpha$ is the spectral index. The detected MRAs are spectrally steep and their fitted spectral indices varied from -3.96 and -5.77.  \citet{var21}, using a sample of 86 MRAs,  measured the spectral indices between -0.650 and -7.106.  The derived spectral indices of 5 MRAs presented in this study clearly fall within the same range. The radio emission is detected simultaneously over the full 30-50 MHz observing band.  The spectra of MRA3 and MRA5 have higher noise compared to the other MRA events.  The gaps in the spectra are due to the 2.6 MHz separation between the observed subbands. The smooth, broadband,
power law spectrum also rules out the possibility of this emission originating  from narrowband, artificial RFI sources.

\begin{table}[h!]
\begin{center}
\begin{tabular}{  c c c c  }
\hline
 Name & Elevation Angle & UT Date & UT Time\\ 
  & (degrees) &  &   \\
  \hline
 MRA1 & 56.067 & 08/10/2018 & 20:49:50 \\
 MRA2 & 69.689 & 08/10/2018 & 21:06:05 \\
 MRA3 & 49.916 & 08/11/2018 & 04:34:09 \\
 MRA4 & 70.787 & 08/11/2018 & 06:38:57 \\
 MRA5 & 45.006 & 08/16/2018 & 13:48:06 \\
 \hline
\end{tabular}
\caption{List of detected MRAs with their corresponding names, elevation angle, UTC date and time.}
\label{mra_table}
\end{center}
\end{table}

\subsection{Higher Angular Resolution Imaging} \label{highim}
Once the MRAs are detected, the next important step is to conduct the higher angular resolution imaging and it consists of four parts. First is the peeling of bright sources, then a near-field correction,  a background image subtraction in the visibility plane to improve the SNR and finally imaging in CASA.  These methods are discussed in the following subsections.

\subsubsection{Peeling of Bright Sources} \label{sec:peel}

  In order to achieve maximum angular resolution, all the baselines are utilized for imaging.  
  However, these images are badly affected by the sidelobe patterns from bright radio sources below 100 MHz such as Cygnus A, Cassiopeia A, Taurus A and Virgo A. These sources are historically referred to as the ``A-team" sources. They have high flux density and they are named on the basis of the constellation of the source followed by the letter `A'.  The strong sidelobes from the A-team sources are primarily due to mutual coupling between nearby antennas and the resultant variation in the beam pattern of individual dipole antennas \citep{mar19}. In order to mitigate this effect, a subtraction of these bright sources and their associated sidelobe patterns from the image is required. This process is called ``peeling" and it is implemented using a custom Python script executed with CASA.  
  
\begin{figure}[ht]
\centering
	\includegraphics[width=0.99\textwidth]{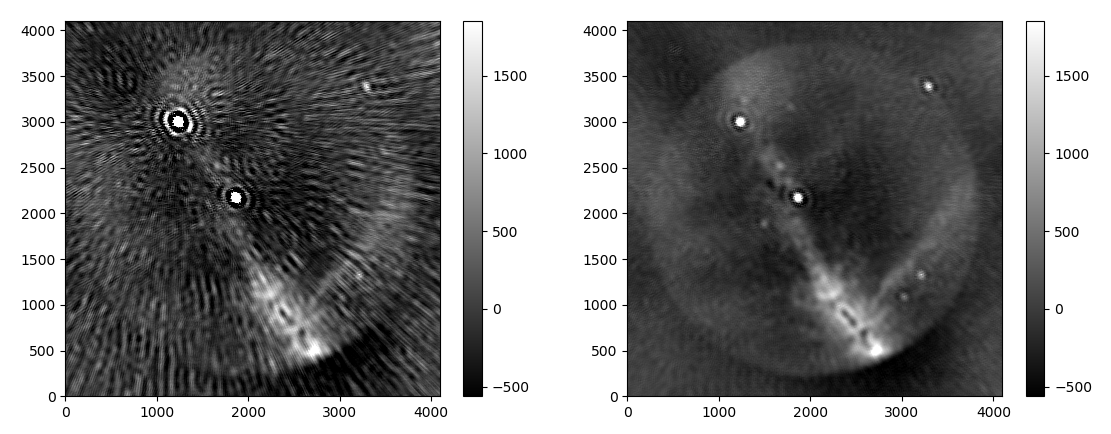}	
	\caption{The OVRO-LWA all-sky image before and after peeling. Left panel shows the image of the sky before peeling showing strong sidelobes from bright sources Cas A (top left) and Cyg A (middle). Right panel shows the image of the sky after peeling with minimized sidelobe patterns from the images. Both images are plotted within the intensity range of the image after peeling for easy comparison. The sources in the image before peeling are 10 times brighter than the sources in the peeled image. The color bar shows the intensity of the pixels in units of Jy/beam }
	\label{peel}
\end{figure}	

In this process, we conduct a direction dependent calibration towards each bright source (in the order of brightness) in an image using a point source model, and the  model is subtracted from the data. This will remove the bright source flux and the associated sidelobes increasing the SNR of the image. 
After the bandpass calibration, a primary imaging with CASA \texttt{clean} task is carried out with 4096 $\times$ 4096 image size, 1.875 arcmin pixel size and Briggs weighting with a robust value of zero. The peak flux density values of each A-team sources as well as their elevations are measured from the image. We peel the A-team sources when they are above the horizon and when their peak flux density is greater than 2000 Jy. Then, the peeling process is carried out in the order of source brightness.  The  sidelobes of sources with flux density less than 2000 Jy are weak and they do not negatively affect the SNR of the image.
Peeling starts by inserting  model visibilities of the source into the measurement sets using the task \texttt{ft}. A phase and an amplitude calibration solutions towards the source direction are derived using task \texttt{gaincal}. The derived solutions are applied to the measurement set using task \texttt{applycal}. The next step is to subtract the model visilbilities from the corrected data using task \texttt{uvsub} and to revert the source specific calibration already applied. This procedure is repeated for each sources in the order of their brightness. After peeling is completed, the measurement sets are imaged again with task \texttt{clean} to obtain an integrated all-sky image over frequency channels in Stokes I.  Figure \ref{peel} shows the difference in all-sky images before and after peeling. The image after peeling has minimized sidelobe pattern from the bright sources Cyg A and Cas A compared to the images before peeling. There are still some residual flux left from bright sources after the model subtraction. The root mean square (RMS) noise of the image after peeling has improved by a factor of two.

\subsubsection{Near field correction} \label{nf}
Generally, in radio astronomy, correlators assume incoming radiation as plane waves,  since astrophysical sources at large distances are in the far field. This plane wave assumption breaks down for nearby sources such as  solar system objects, comets, etc, when they are relatively in a near-field with respect to the interferometer. The condition for near-field to far-field transition is given in equation 1 and discussed in section 4. When the sources are within the near-field, a correction to account for the curvature of incoming wave needs to carried out before imaging. In the visibility (correlated voltage data product for each pair of antenna) space, this can be achieved by multiplying the visibilities of each baseline with a phase factor given by
\begin{equation}
    F = e^{i2\pi\delta\nu}
\end{equation}
where $\delta$ is the delay in the arrival of signals between a pair of antenna and $\nu$ is the frequency of observation. In the case of plane wave approximation, $\delta$ is the projection of a baseline towards the source direction, usually known as the geometric delay. However, when the source is within the near-field, especially for MRAs, we calculate the signal arrival time for each antenna in the array. From the transient search pipeline, we know the azimuth and elevation of the MRA detection and here we make an assumption that the MRA occurred at a distance of 100 km from the center of the array.  
Most of the MRAs are observed between 90-120 km altitudes \citep{var19b, ob16b}. Even though that is the case, different parts of the meteor plasma trail could be at different altitudes. Since we lack observations from other LWA stations or optical cameras, an accurate estimation of the distance to each part of the trail is not possible. The distance of 100 km was chosen as a nominal distance to which we can apply the near-field correction.
The azimuth, elevation angle and distance of the MRA are converted to Earth Centered, Earth Fixed (ECEF) {\tt X,Y,Z} coordinates and the distance between the MRA and each antenna stand (in ECEF {\tt X,Y,Z} coordinates) is calculated. This distance is divided by the speed of light to get the signal arrival time for each antenna. Then,  the difference in the signal arrival time for each antenna pair ($\delta$) is used to calculate the phase factor which is applied to the visibilities for each antenna pair. This will result an image of the sky phased towards the MRA and corrected for the near-field.

\subsubsection{Visibility Subtraction and Imaging} \label{vissub}
In the transient search pipeline, we utilize image based subtraction to reduce the contributions of steady sources in the sky. This method is usually efficient and faster when we have to process large amounts of data, but, at some level,  the image plane is still affected by side lobes and structures associated with the bright sources. A subtraction of the background sky from the peak MRA sky in the visibility plane can be more effective in reducing the effects of side lobes associated with steady sources. The visibility subtracted MS data set was imaged using CASA. For imaging, \texttt{clean} task with 4096 $\times$ 4096 image size, 1.875 arcminute pixel size, Briggs weighting with a robust value of zero and an interactive deconvolution was used. In the interactive deconvolution procedure, the whole field of view was cleaned initially with 500 iterations and later, deconvolving boxes were used around the sources of interest to clean specifically in multiple iterations. This will remove the point spread function or the associated dirty beam effects in the region of interest.

\section{Results}
\subsection{Spatially Resolved Observations of MRAs} \label{results}
The higher angular resolution imaging resulted in the spatially resolved observations of 5 MRAs. Figures 4-8 shows the resolved observations of all MRAs in Stokes I and their evolution in time and frequency. The corresponding synthesized beam in each image is shown as an ellipse at the left bottom of each subplot. The synthesized beam becomes smaller as the frequency increases. Even though we should observe the same synthesized beam size in all MRAs at the same frequency subbands, the flagging of some long baselines due to technical and RFI issues can cause the beam sizes to vary. This is also a function of elevation angle since that projects down the array. This has resulted in slightly different sized ellipses in each MRA plots. With  the 1.5 km baseline at 40 MHz, we should achieve $\sim$18 arcminute resolution.

MRA1 has a total duration of 39 seconds and an elongated source structure with an angular extent of $\sim$ 7 degrees which scales to $\sim$ 14 km lengths at 100 km height. The emission is bright in the first 13 seconds and the components resolve out towards higher frequencies as expected.  In the second integration, the emission is weak in the highest subband. The third integration has a weak emission in the first low frequency subband and it is absent in the high frequency subbands. 

MRA2 is a 65 seconds long duration event and has an elongated trail with an angular extent of $\sim$ 16 degrees which scales to $\sim$ 30 km lengths at 100 km height. There is a position shift in the emission regions as a function of time. In the first integration, all the bright components of the trail resolves out well towards higher frequencies. The length of the trail becomes shorter after the first integration. The radio emitting hot spots move slowly in the direction of the meteor entry and slowly confine to a smaller region towards the last integration. The radio emission in the 65 and 78 s integration decays faster towards higher frequencies. Also, there are some notable negative holes in the images from 26-52 second integrations. These are due the scintillation of a bright VLA Low Frequency Survey (VLSS:\citet{Coh07}, \citet{lan12}) source, with a flux density of 133 Jy at 74 MHz. The source brightness is changing as a function of time and the higher source flux in the noise image is over subtracted from the MRA image. 

MRA3 is a 13 s event with an elongated source structure making an angular extent of $\sim$ 27 degrees which scales to $\sim$ 60 km lengths at 100 km height. This is the longest MRA trail detected in this search.  Even though  the brighter components get resolved towards the high frequencies, they are weak.

MRA4 is a 39 s duration event with mostly point source structure. The third subband at 41.760 MHz is completely affected by RFI and removed from the analysis. The images are also affected by the scintillation of bright radio sources even after peeling and there are many noise structures around the MRA hot spots. The emission is relatively strong and unresolved across time and frequency.

MRA5 is a 13 s event with point source structure and resolved at high frequencies. In the last subband, the source is resolved into 3 main components. Figure \ref{mra5_comp} shows the last subband (46.992 MHz) image of MRA5 using the core array at lower resolution and with the full array at the higher resolution. The MRA is a point source at lower resolution, and can be clearly seen as 3 distinct components in the higher resolution image. Each component is resolved down to less than the synthesized beam width of the image.

Full Stokes I, Q, U and V imaging of the MRAs was also conducted to understand the polarization of the emission regions. This was only conducted for the first integration and last subband dataset at 46.992 MHz where MRAs are bright and well resolved. Figure \ref{mra2_pol} shows the full Stokes parameter images of MRA2.  The MRA components are resolved down to the synthesized beam in Stokes I and there are several smaller point sources smaller than the beam size along the trail. However, the emission is absent in Stokes Q and weak in Stokes U and V (less than 20 \%). The weak emission levels in Stokes U and V can be attributed to the expected polarization leakage of LWA antennas \citep{ob15a}. \citet{east18} also measured the Stokes Q polarization leakage values for OVRO-LWA and it agrees with the LWA1 measurements reported by \citet{ob15a}. Similar full Stokes parameter imaging of other MRAs also showed lack of polarized emission. This suggests that the linear and circular polarization is heavily suppressed in MRAs at these spatial resolutions. 
\begin{figure}[h!]
	\centering
	\includegraphics[height = 9 cm, width = \textwidth]{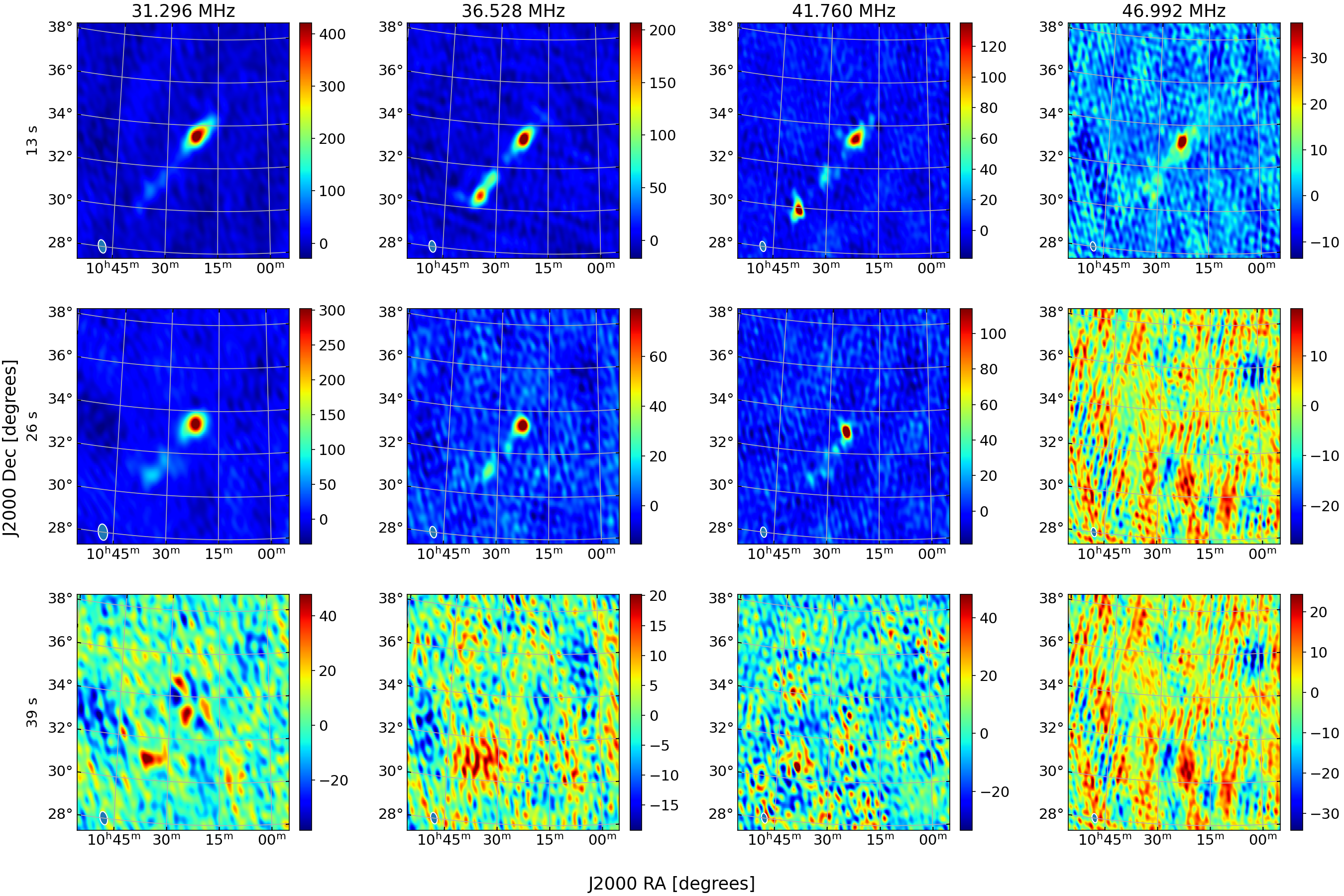}	
	\caption{The plot shows the evolution of MRA1 as a function of frequency subband and time integrations. Each row represent a time integration which is marked on the left side of each row. Each column represent a subband and the corresponding center frequency is marked on the top of the column. The color bars of each subplot are in sky intensity values in Jy/beam. The synthesized beam of the telescope is shown as an ellipse at the left bottom corner of each subplot. Each subplot is given a different color bar to reveal the faint emission structures.}
	\label{grid_mra1}
\end{figure}	

\begin{figure}[h!]
	\centering
	\includegraphics[width = \textwidth]{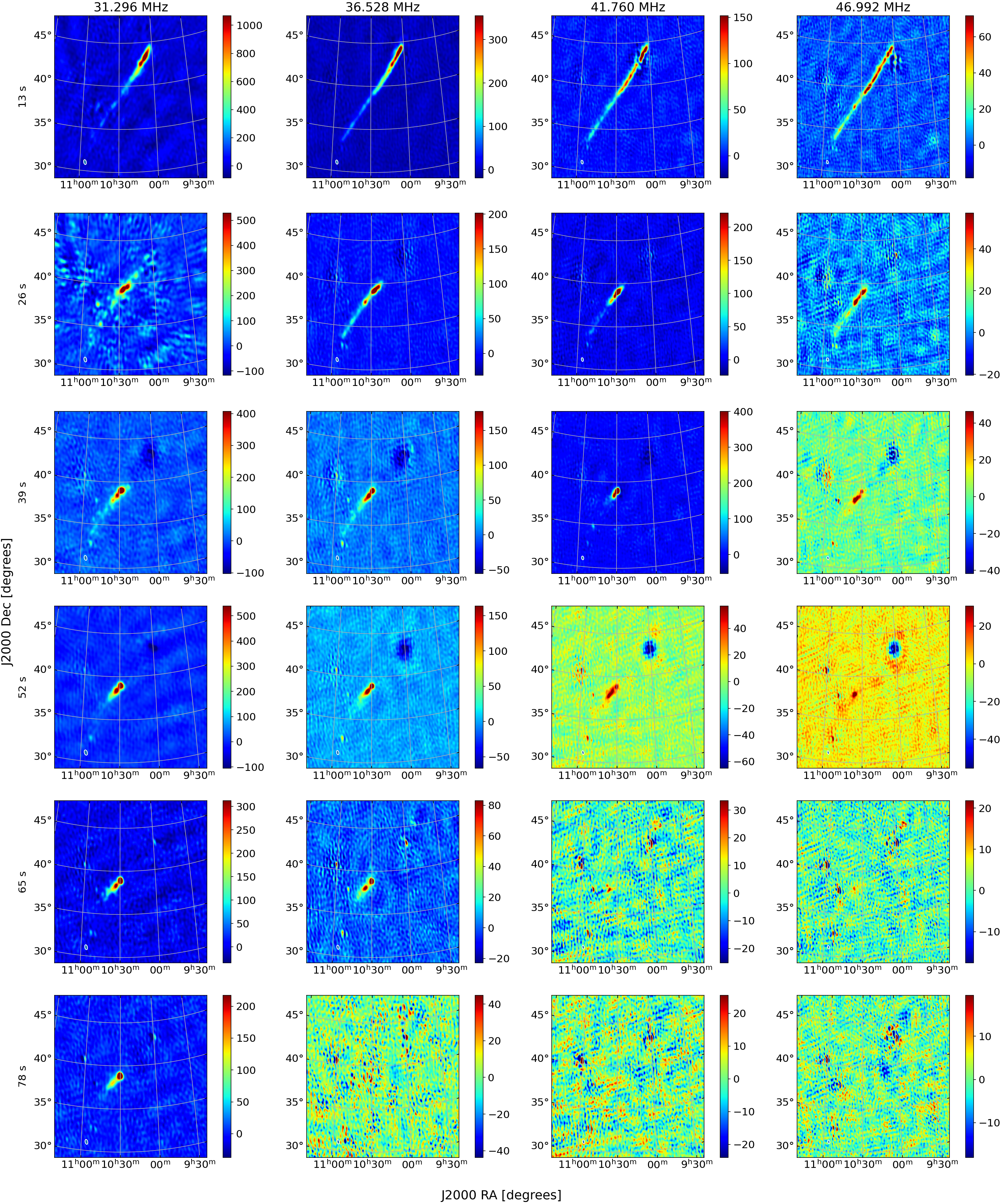}	
	\caption{The plot shows the evolution of MRA2 as a function of frequency subband and time integrations. The layout and labeling are the same as Figure \ref{grid_mra1}. }
	\label{grid_mra2}
\end{figure}

\begin{figure}[h!]
	\centering
	\includegraphics[height = 3.5cm, width = \textwidth]{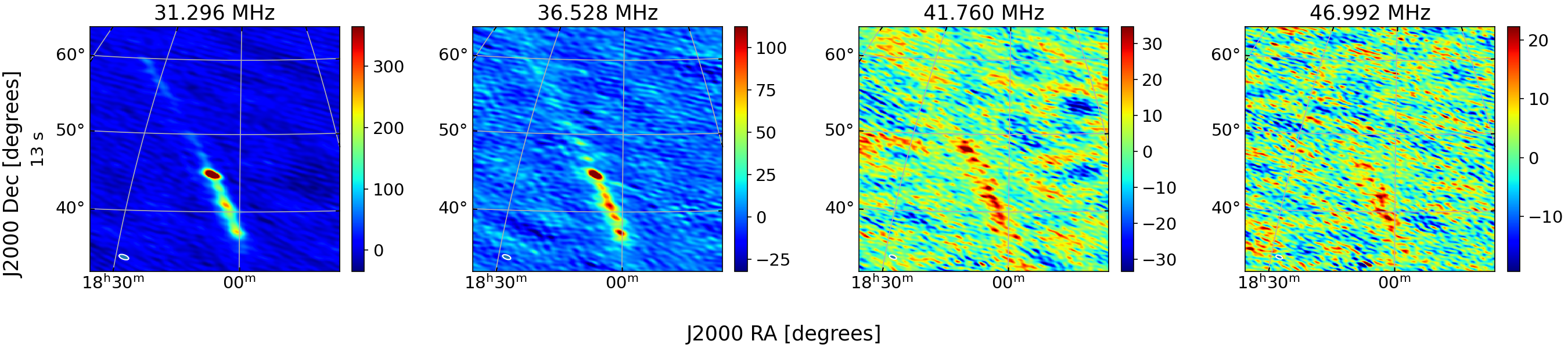}	
	\caption{The plot shows the evolution of MRA3 as a function of frequency subband and time integrations. The layout and labeling are the same as Figure \ref{grid_mra1}. }
	\label{grid_mra3}
\end{figure}	

\begin{figure}[h!]
	\centering
	\includegraphics[width = \textwidth]{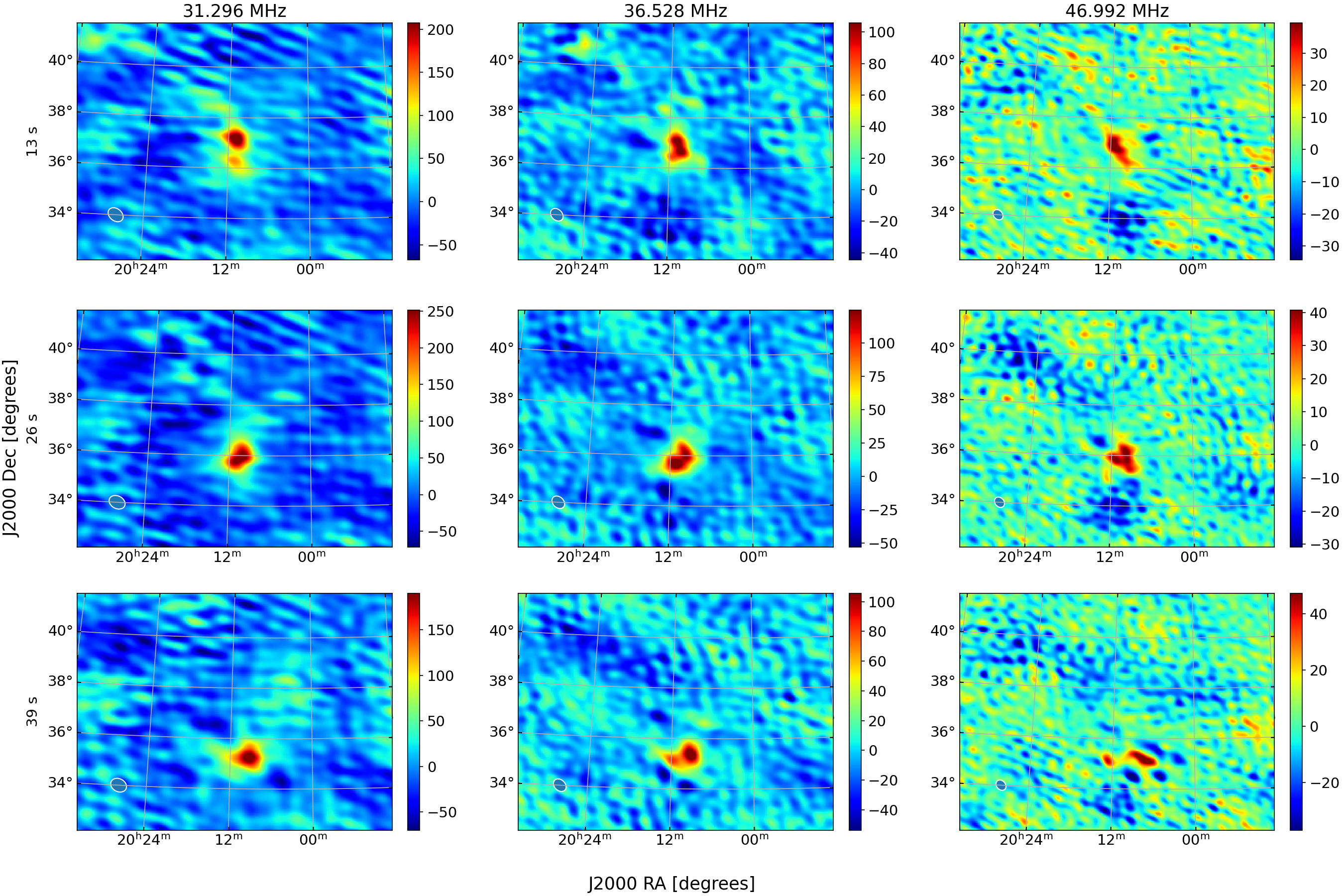}	
	\caption{The plot shows the evolution of MRA4 as a function of frequency subband and time integrations. The layout and labeling are the same as Figure \ref{grid_mra1}. }
	\label{grid_mra4}
\end{figure}	

\begin{figure}[h!]
	\centering
	\includegraphics[height = 3.8cm, width = \textwidth]{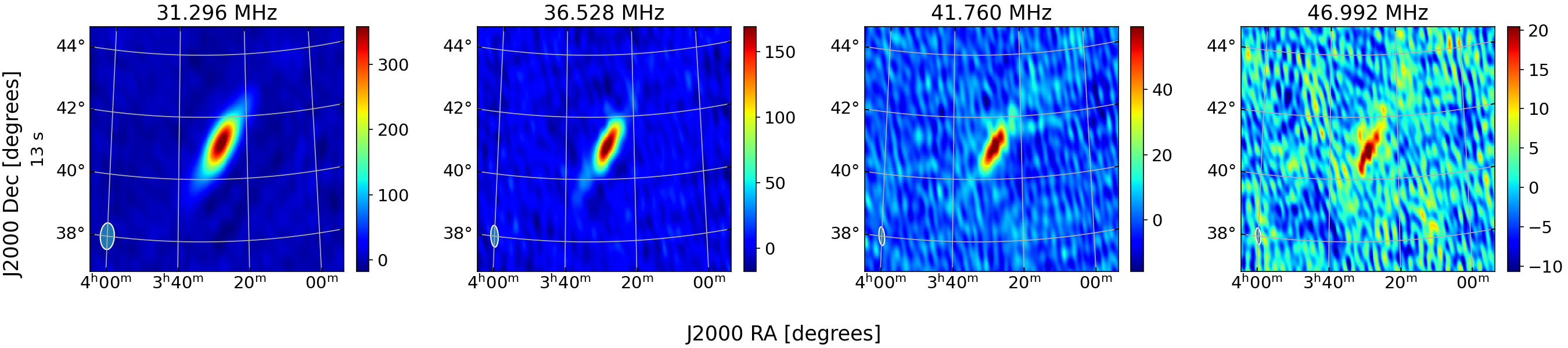}	
	\caption{The plot shows the evolution of MRA5 as a function of frequency subband and time integrations. The layout and labeling are the same as Figure \ref{grid_mra1}. }
	\label{grid_mra5}
\end{figure}	

\begin{figure}[h!]
	\centering
	\includegraphics[ width = \textwidth]{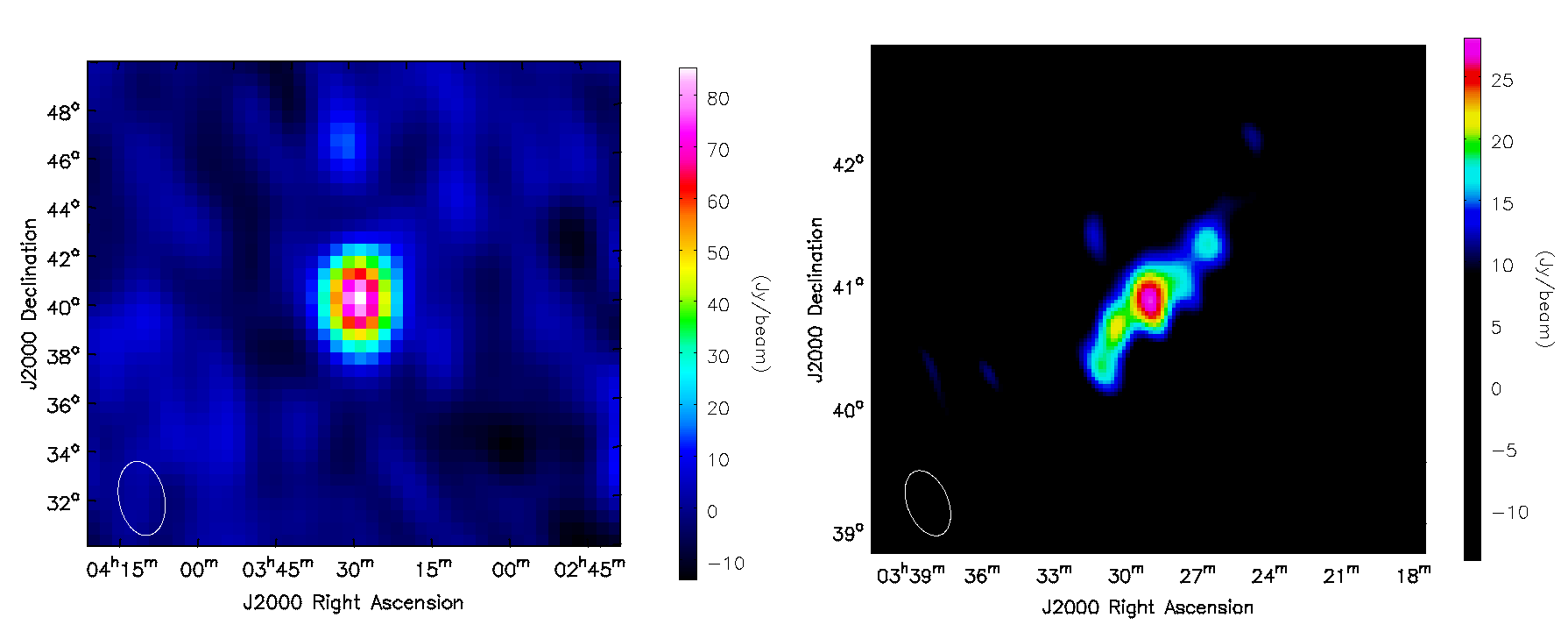}	
	\caption{Plot showing the last subband  Stokes I image of MRA5 (46.992 MHz) at lower angular resolution (left) and higher angular resolution (right). The color mapping is adjusted to suppress the noise in the image.}
	\label{mra5_comp}
\end{figure}	

\begin{figure}[h!]
	\centering
	\includegraphics[ width = \textwidth]{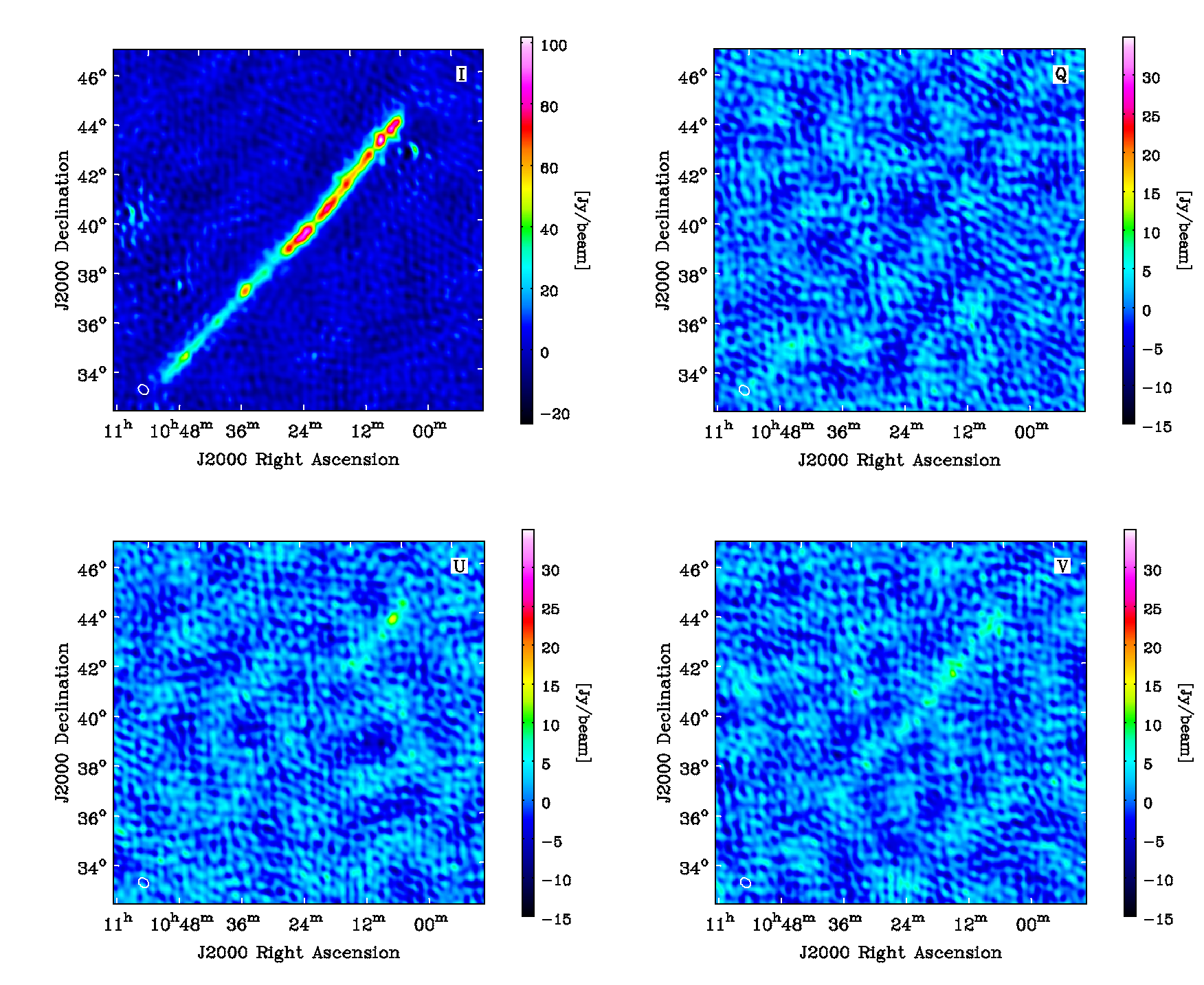}	
	\caption{Stokes I,Q,U and V images of MRA2 in the last subband (46.992 MHz) and the first integration (13 s) where the emission is brighter, resolved and longer compared to other integrations.}
	\label{mra2_pol}
\end{figure}	
\subsection{Spectral Index Mapping} \label{simap}
In order to study the energy distribution of the different resolved components in the higher angular resolution images, a spectral index mapping was created for all MRAs. For that, first we need to ensure that the images from each subband have the same synthesized beam. 
We estimate the size of the lowest frequency beam using \texttt{ia.restoringbeam} and then use \texttt{ia.convolve2d} in CASA to match the other frequencies to that.
Then, a Python script is used to fit a power law for each pixel across frequencies and their spectral index (SI) value is calculated.  This process is carried out for each pixel with intensity values greater than 3.5 times the standard deviation of the off-source region in the intensity map. The pixels not fulfilling this threshold criteria and the ones that failed the power law fitting are masked in the SI maps. This step to is carried out to mitigate spectral index values with high uncertainties resulting from the poor power law fitting of noisy components.

\begin{figure}[h!]
	\centering
	\includegraphics[height = 8cm, width = \textwidth]{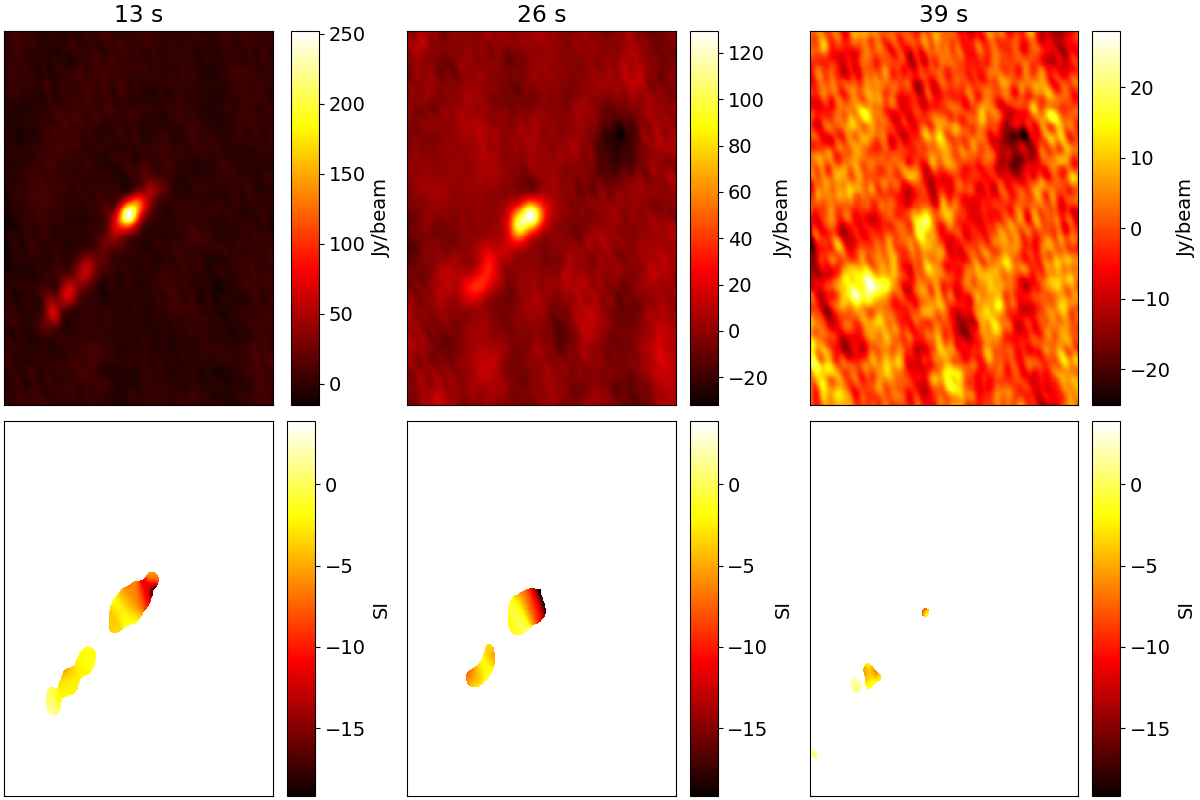}	
	\caption{Plot showing the averaged Stokes I intensity and spectral index maps of MRA1. Top panel shows intensity maps averaged over 4 subbands. Each column represent a time integration. The bottom panels shows the corresponding spectral index maps created for each time integration.}
	\label{si1}
\end{figure}

\begin{figure}[h!]
	\centering
	\includegraphics[height = 16cm, width = \textwidth]{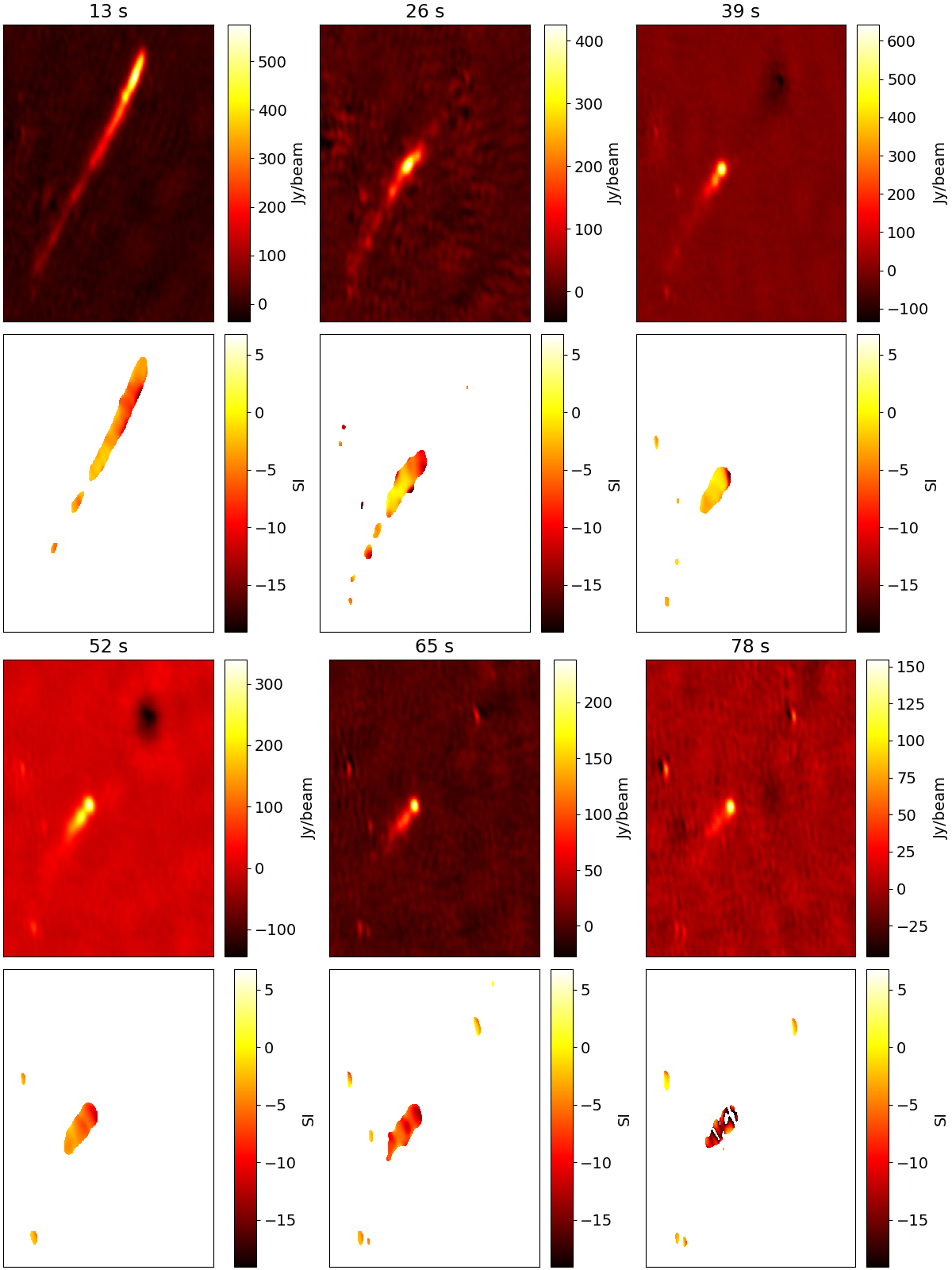}	
	\caption{Plot showing the averaged Stokes I intensity and spectral index maps of MRA2. The layout and labeling are the same as in Figure \ref{si1}. The first two rows corresponds to the intensity and SI maps from 13 s to 39 s. Similarly third and fourth row shows 52-78 s. }
	\label{si2a}
\end{figure}

\begin{figure}[h!]
	\centering
	\includegraphics[height = 8cm, width = 5cm]{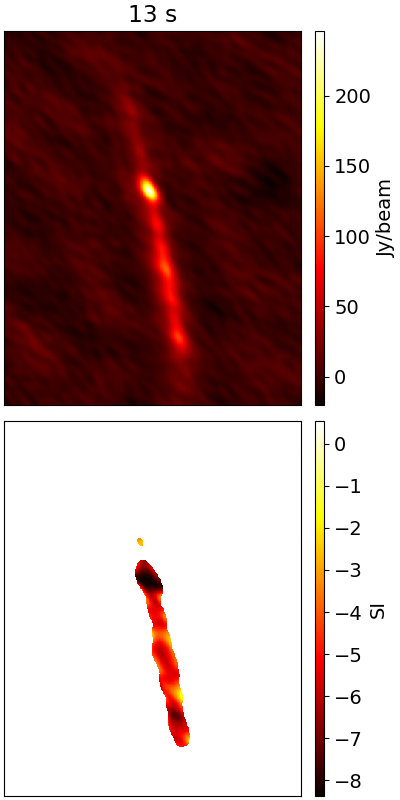}	
	\caption{Plot showing the averaged Stokes I intensity and spectral index maps of MRA3. The layout and labeling are the same as in Figure \ref{si1}.}
	\label{si3}
\end{figure}	

\begin{figure}[h!]
	\centering
	\includegraphics[height = 8cm, width = \textwidth]{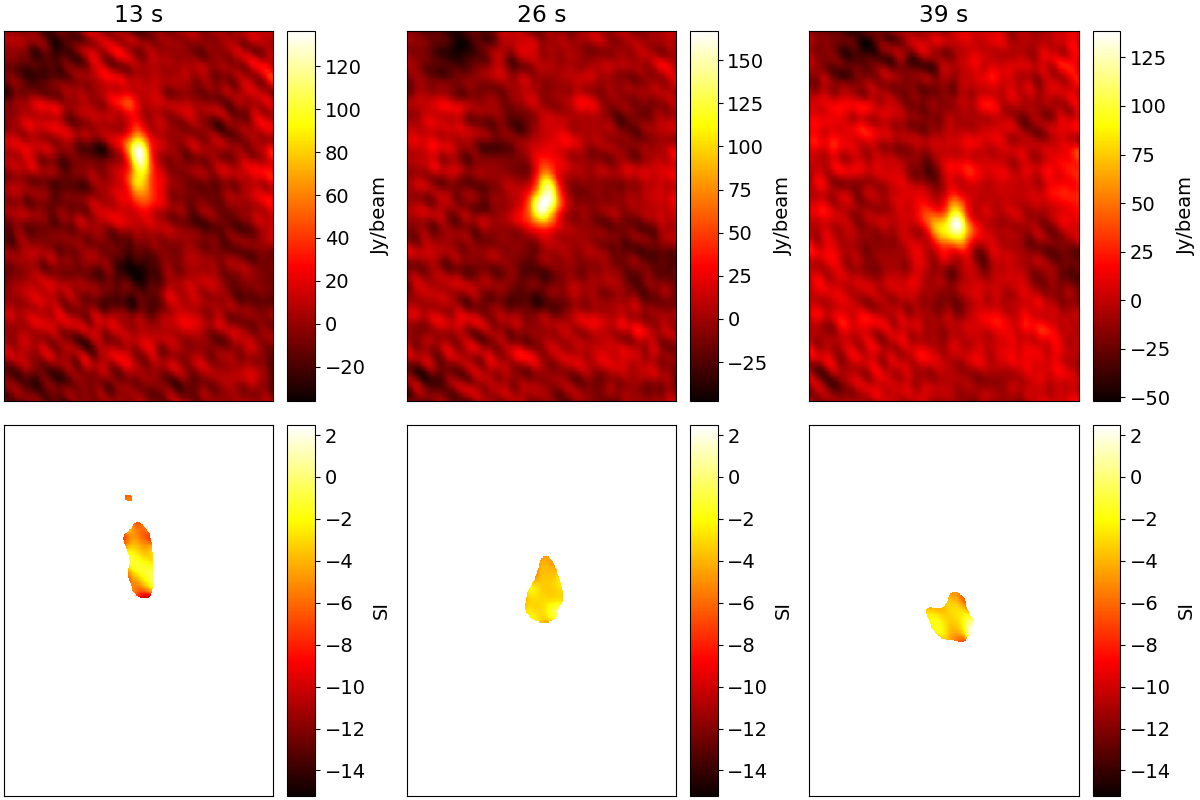}	
	\caption{Plot showing the averaged Stokes I intensity and spectral index maps of MRA4. The layout and labeling are the same as in Figure \ref{si1}.}
	\label{si4}
\end{figure}	

\begin{figure}[h!]
	\centering
	\includegraphics[height = 8cm, width = 5cm]{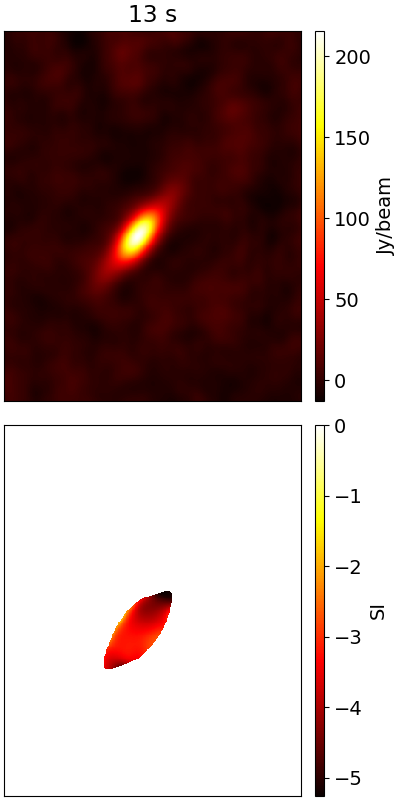}	
	\caption{Plot showing the averaged Stokes I intensity and spectral index maps of MRA5. The layout and labeling are the same as in Figure \ref{si1}.}
	\label{si5}
\end{figure}	

Figures 11-15 shows the Stokes I intensity maps averaged over the subbands and their corresponding spectral index maps for each integration. The SI maps show the evolution of the spectra of each component with time. The positive or close to zero SI values indicates flat spectra where the emission is more or less the same across all frequencies. The negative SI values indicates steep spectra where emission flux is higher at lower frequencies and lower at high frequencies. The  more negative the SI values, the steeper the spectra are.

The SI map of MRA1 is a combination of flat and steep components in the first 2 integrations and the emission recedes in the final integration. The SI map of MRA2 is a combination of flat and steep components from 0-39 s. However, the emission components get steeper over time from 52-78 s. The emission flux in the last integration is completely dominated by components that are steeper than the previous integrations. 


The SI map of MRA3 is a combination of flat and steep components, where most of the emission is dominated by the steep components.  In contrast, the SI map of MRA4 is mostly dominated by flat components and the emission flux is more or less consistent across time. Similar to MRA3, the SI map of MRA5 is also a combination of flat and steep components, where the SI distribution is dominated by the steep components.

\section{Discussion}

We expect that MRA producing meteor trails consist of turbulent plasma, where the electron density varies significantly over even small size scales. \citet{Oppen15} has shown that the neutral winds cause meteor trails to quickly become turbulent after formation. We would expect the level of turbulence to be even greater for large MRA producing meteors as they are likely turbulent when they form due to a high Reynolds number from a large ($>$ 1 cm) object traveling at 60 km/s. Uneven ablation, fragmentation and changing wind velocities with altitude would introduce variability in turbulence along the trail. The MRAs presented here (especially MRAs 1, 2, and 3) clearly show variation in brightness that could be interpreted in terms of  variation of turbulence. Under the RTR hypothesis, the luminosity of the emission depends on the level of turbulence. The luminosity would also depend on the number of hot electrons present. As described in \citet{ob20}, these hot electrons could be created by the oxidation of anions. The variation in luminosity of the MRAs presented in this paper, is consistent with the level of variation observed in typical persistent trains, which are tracers of the chemical reactivity in meteor trails \citep{ob20}. The levels of turbulence and reactivity could be linked, where increased turbulence and mixing with the ambient atmosphere could increase the reaction rate.

The higher angular resolution imaging has spatially resolved MRA emission down to less than 30 arcminutes (major axis of the synthesized beam) at 47 MHz. This translates to a physical scale of 1 km for an MRA observed at a zenith angle of 30 degrees and at 100 km altitude. Structured brightness variations are clearly visible, implying that the MRA emission consists of several small scale plasma structures of the order of 1 km or less. This is 8-10 times smaller than the plasma size scales reported by previous LWA observations \citep{ob14b}. The earlier work of \citet{var19b} based on the isotropic radiation pattern of MRAs proposed that the emission has to be either incoherent on large plasma scales ($\sim$10-12 km) or an incoherent addition of several small scale coherent structures. 
Coherent emission would be both directional and polarized. In this work, we have found no evidence that such processes are occuring at size scales of at least 1 km.

It is possible that the emission could be either completely incoherent at 1 km size scales or it could be an incoherent addition of further smaller coherent structures ($\ll$1 km). This would require further observations with longer baselines than the stage II of OVRO-LWA. However, there is no evidence of any coherent process occuring and it is doubtful that increasing the resolution again by 10 fold would find any.

The SI maps of MRA2 clearly demonstrates the diffusion of the plasma structures over time. This diffusion effect was also observed in the spectral work carried out by \citet{ob15b, ob16a}. In the initial half of the duration of MRA2, the emission is dominated by flat spectra components. In the later half, the emission is dominated by steep spectra components. 
Under the RTR hypothesis, the MRA producing turbulent plasma emits at the local range of plasma frequencies.
Then, the radio spectral index at different times is likely related to the distribution of plasma, where a flat radio spectra indicates a equal amount of plasma at different plasma frequencies. Similarly, a steep spectra indicates more plasma at lower plasma frequencies. 

During the initial meteor entry, the  trail mostly consists of high density plasma, which quickly becomes turbulent. Over the course of time, the turbulent trail diffuses out to the ambient medium, reaching equilibrium with the background low density plasma. During diffusion, there would be an increasing amount of low density plasma emitting at lower frequencies compared to the small amount of high density plasma emitting at high frequencies, resulting in a steeper spectra. This effect of diffusion can be clearly seen in MRA2. 
However, this effect is not seen in other MRAs. MRA3 and MRA5 are single integration events and this effect cannot be observed.  MRA1 and MRA3 are 39 second duration events and the effect of diffusion is not observed within this time scale. From the properties of MRA2, the radio emission needs to be brighter across frequencies and longer duration than a minute to observe diffusion effects.

The Langmuir wave conversion is a coherent process which subsequently needs to directional and polarized. However, the RTR emission can be incoherent and unpolarized. From Figure \ref{mra2_pol}, there are many point-like emission regions smaller than the synthesized beam which are unpolarized. This suggest that at a physical size scale of the order of 1 km, the emission is still incoherent. Therefore, the current evidences suggest that the RTR hypothesis might better explain the emission mechanism of MRAs compared to Langmuir wave conversion. 
Even though a lack of polarization was observed in all of the spatially resolved MRAs presented in this paper, future observations of a large sample of MRAs are required to substantiate this finding.  Also, it is worth noting that our resolution ($\sim$ 1 km) is only 100 times greater than the observed electromagnetic wavelengths ($\sim$ 10 m).

The multi-instrument observations of the Pajala fireball conducted by \citet{vier22} shares some interesting properties with the MRAs presented in this paper. Radar observations of a fireball induced non-specular meteor trail shows similar temporal behavior of MRAs in this work. Certain altitudes showed a longer duration echo and the trail was observed to deform due to the neutral winds. A similar effect has been observed for the resolved radio structures of the long duration events, MRA1 and MRA2 (predominantly). The radar echoes reported in \citet{vier22} also display varied onset times at different altitudes. The authors suggest this is due to variability in plasma turbulence formation. These altitude/onset effects are also observed in the resolved MRA trails presented in this paper, and since the RTR mechanism requires a significant amount of turbulence we would expect to see a connection between MRAs and non-specular echoes. \citet{var21} has already mentioned the possible connection between MRAs, persistent trains and long duration non-specular echoes caused by charged dust and turbulence. The similarities observed between MRAs and long duration non-specular echoes provides further evidence to support this connection. A simultaneous observations of range spread non-specular meteor trail echoes and MRAs in the future may shed more light into the nature of this phenomena.

\section{Conclusions  and Future Work}
In this work, we conducted transient searches focused on the detection of MRAs with the OVRO-LWA data collected during the Perseids meteor shower in 2018. This resulted in the detection of 5 MRAs. This is the first time MRAs have been detected with the OVRO-LWA station. The data collected is used to conduct higher angular resolution observations and probe the emission size scale of MRAs.  This resulted in the spatially resolved observations of MRAs down to less than 30 arcminutes and this translates to less than 1 km physical lengths at 100 km heights. The intensity maps of the resolved MRAs clearly shows the evolution of radio trails across frequency and time. The spectral index maps of MRAs revealed the diffusion of meteor plasma in MRA2 which is a 65 s long duration event.  The lack of polarization in the resolved MRA emission suggests that the emission is still incoherent on 1 km size scales or it could be an incoherent addition of further smaller coherent regions. However, verifying this requires observing with the stage III of the OVRO-LWA which can provide baseline lengths up to 2.6 km. 
The lack of polarization in the resolved spatial structures of all 5 MRAs down to less than 1 km physical size scales at 100 km heights favour RTR hypothesis over the Langmuir hypothesis to explain the radiation mechanism of MRAs. However, more work in the future is
required to improve the understanding on the accurate emission size scales of MRAs and to understand the emission mechanism.

For the future work, the stage III of the OVRO-LWA can offer up to 2.6 km baseline lengths resulting an angular resolution of 9 arcminutes at 40 MHz, allowing us to probe plasma scale structures on the order of 300 m.  However, detection might be difficult at these spatial scales depending on the properties of the emission mechanism. The best approach would be similar to this work, in which we utilize the core array for detection and use the long baselines for getting a better constraints on the emission size scales. Also, the recently developed fast visibility mode for the stage III, providing 100 ms integration visibilites from 48 dipole stands of the array would be interesting to study the dynamics of the MRA.  Even though the spatial resolution would go down, the fast cadence observations will be able to map out the effects of neutral winds on MRAs. 

Another future project would be to use an optical telescope to probe persistent trains down to size scales much smaller than achievable by the LWA stations used to date. The upcoming LWA Swarm, consisting of multiple LWA stations could achieve 15 m spatial resolution. A modest 85 mm lens on a DLSR for instance could achieve 15 m resolution. From here, we could get additional turbulence parameters which we could then compare to MRAs. One of the authors in this work, Kenneth Obenberger at AFRL is currently working on this project to get two all sky cameras to trigger a robotic azimuth/elevation mount with a Sony camera equipped with a 85 mm f/1.8 lens.

\acknowledgments
Support for operations and continuing development of the LWA1 is provided by the Air Force Research Laboratory and the National Science Foundation under grants AST-1711164, AST-1835400 and AGS-1708855. This work utilized the above grants to conduct this research. We would like to thank the UNM Center for Advanced Research Computing, supported in part by the National Science Foundation, for providing the research computing resources used in this work. The data used in this work is based in part upon work supported by the National Science Foundation under grant number AST-1828784, the Simons Foundation (668346, JPG), the Wilf Family Foundation and Mt. Cuba Astronomical Foundation.  

\section*{Open Research}
\subsection*{Data Availability Statement}
The calibrated data of 5 MRAs \citep{OVRO18} in the CASA measurement set format were used in the preparation of this manuscript.  The CASA software \cite{mc07} can be used to read the measurement set data and image the MRAs.

\bibliography{agusample}

\end{document}